\documentclass[sigconf]{acmart}

\usepackage{xcolor}
\usepackage{framed}
\definecolor{shadecolor}{gray}{0.93}

\makeatletter
\renewenvironment{shaded}
{%
  \def\FrameCommand##1{%
    \fboxrule=0.8pt\relax
    \fboxsep=8pt\relax
    \fcolorbox{black}{shadecolor}{##1}%
  }%
  \MakeFramed{\advance\hsize-\width \FrameRestore}%
}
{%
  \endMakeFramed
}
\makeatother

\usepackage{subcaption}
\usepackage{listings}
\usepackage{dblfloatfix}
\AtBeginDocument{%
  }

\copyrightyear{2026}
\acmYear{2026}
\setcopyright{cc}
\setcctype{by-nc-nd}
\acmConference[CHI '26]{Proceedings of the 2026 CHI Conference on Human Factors in Computing Systems}{April 13--17, 2026}{Barcelona, Spain}
\acmBooktitle{Proceedings of the 2026 CHI Conference on Human Factors in Computing Systems (CHI '26), April 13--17, 2026, Barcelona, Spain}
\acmPrice{}
\acmDOI{10.1145/3772318.3791567}
\acmISBN{979-8-4007-2278-3/2026/04}

\begin{document}

\title[Improving UI Generation Models from Designer Feedback]{Improving User Interface Generation Models from Designer Feedback}

\author{Jason Wu}
\authornote{Work performed while author was affiliated with Apple}
\affiliation{%
  \institution{Purdue University}
  \city{West Lafayette}
  \country{USA}}
\email{jasonwu@purdue.edu}

\author{Amanda Swearngin}
\affiliation{%
  \institution{Apple}
  \city{Seattle}
  \country{USA}}
\email{aswearngin@apple.com}

\author{Arun Krishna Vajjala}
\affiliation{%
  \institution{Apple}
  \city{Seattle}
  \country{USA}}
\email{a_krishnavajjala@apple.com}

\author{Alan Leung}
\affiliation{%
  \institution{Apple}
  \city{Seattle}
  \country{USA}}
\email{alleu@apple.com}

\author{Jeffrey Nichols}
\affiliation{%
  \institution{Apple}
  \city{Seattle}
  \country{USA}}
\email{jwnichols@apple.com}

\author{Titus Barik}
\affiliation{%
  \institution{Apple}
  \city{Seattle}
  \country{USA}}
\email{tbarik@apple.com}

\renewcommand{\shortauthors}{Wu. et al.}

\begin{abstract}
Despite being trained on vast amounts of data, most LLMs are unable to reliably generate well-designed UIs. Designer feedback is essential to improving performance on UI generation; however, we find that existing RLHF methods based on ratings or rankings are not well-aligned with with designers' workflows and ignore the rich rationale used to critique and improve UI designs. In this paper, we investigate several approaches for designers to give feedback to UI generation models, using familiar interactions such as commenting, sketching and direct manipulation. We first perform an evaluation with 21 designers where they gave feedback using these interactions, which resulted in \textasciitilde1500 design annotations. We then use this data to finetune a series of LLMs to generate higher quality UIs. Finally, we evaluate these models with human judges, and we find that our designer-aligned approaches outperform models trained with traditional ranking feedback and all tested baselines, including GPT-5.

\end{abstract}



\begin{teaserfigure}
  \includegraphics[width=\textwidth]{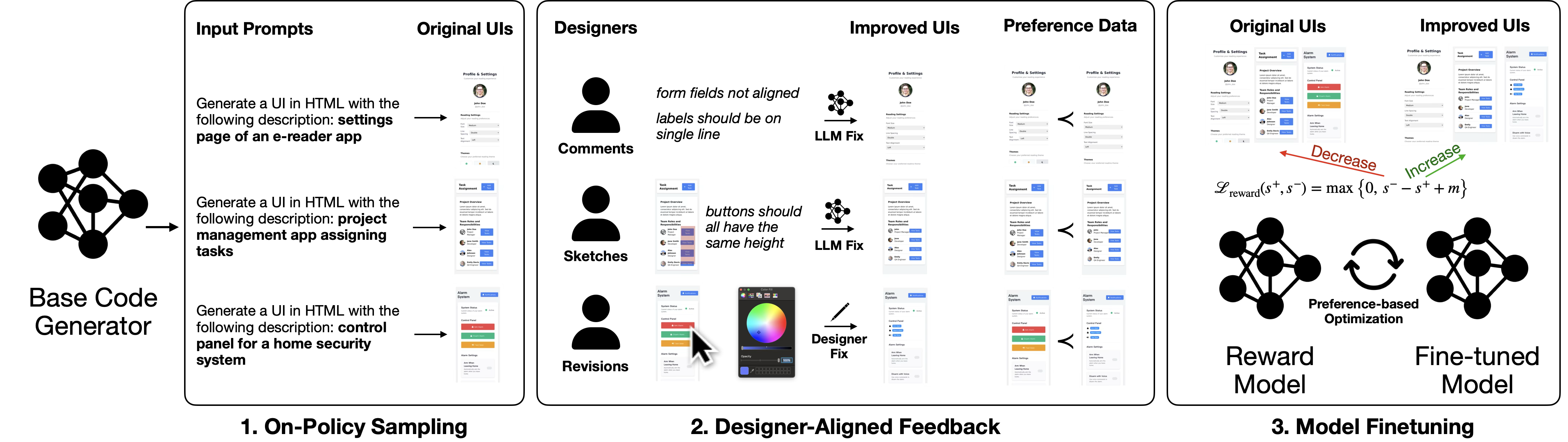}
  \caption{An overview of our approach for improving UI generation models from designer feedback. \textmd{First, we conducted on-policy sampling by generating UI samples using a code LLM. Next, designers employed familiar workflows such as commenting, sketching, and direct revisions to fix design flaws and make improvements. Using the resulting preference dataset, we fine-tuned the code LLM to increase the relative probability of generating code that resembles the improved UIs over code that resembles the original UIs.}}
  \Description{A three pane diagram that shows the three steps to our model fine-tuning approach. In the first pane, a code generator is prompted with UI descriptions which results in rendered UI screenshots. In the second pane, designers give feedback on these screenshots via commenting, sketches, and revisions. Examples of each type of feedback are shown, such as comments that state ``form fields are not aligned.'' These pieces of feedback are used by LLMs and designers to improve the UI screen. In the third pane, the code generator model is trained to generate screens that look more like the designer-improved UIs over the original UIs.}
  \label{fig:teaser}
\end{teaserfigure}

\maketitle

\section{Introduction}
Despite being trained on vast amounts of data, today's large language models (LLMs) are unable to reliably generate well-designed user interfaces (UIs)~\cite{wu2024uicoder}.
This suggests that current datasets do not capture what constitutes ``good'' UI design, much of which resides in tacit domain knowledge~\cite{son2024demystifying}.
Although general heuristics exist~\cite{shneiderman2016eightgoldenrules,nielsen1990heuristic}, applying them effectively requires experience to understand and navigate subtle trade-offs.
As both prospective users and domain experts, designers are well positioned to inform the training of UI generation models.
How can models effectively learn from designers' expertise?

Machine-learning approaches, such as reinforcement learning from human feedback (RLHF)~\cite{ouyang2022training}, have been developed to steer models toward human-preferred responses; however, collecting data required to represent design knowledge is difficult.
For example, prior work applied rubric-guided ratings and rankings to collect designer feedback~\cite{duan2024uicrit,luther2014crowdcrit} but found that subjectivity within the guidelines and differing preferences among designers led to a noisy learning signal for model training~\cite{duan2024uicrit,wu2024uiclip}.
Moreover, these interfaces elicit judgments about existing UIs rather than the changes designers would make, such as concrete edits to layout, colors, and typography.

To understand which types of data are most effective for training models, we investigate designers' common workflows.
For example, previous work~\cite{hartmann2010d} has shown that designers are already effective at communicating design knowledge as part of their day-to-day job through activities such as design reviews, white-boarding, and using design software (e.g., Figma, Sketch). 
These activities produce artifacts such as i) natural language comments, ii) visually-grounded annotations, and iii) low-level design revisions (e.g., before and after a designer's edits) that might be useful for model training. 
To this end, we develop interfaces for model training based on these activities and introduce techniques that transform them into machine-learnable preference data for UI generation models.

To validate our approach, we performed a designer feedback study with 21 professional designers at a large technology company, which led to a dataset of 1460 training examples.
To measure the quality of this generated data, we analyzed this data and found that data collected from designer-aligned interfaces led to higher agreement rates than data collected from conventional ranking interfaces.
To demonstrate the utility of our dataset, we fine-tuned several existing code generation models.
First, we found that supervision derived from designers' sketch and revision feedback yielded significant gains over both untuned baselines and models tuned on conventional ranking feedback.
Second, applying designer supervision to a strong open-source model enables it to outperform all tested baselines, including a larger, state-of-the-art, proprietary reasoning model, GPT-5.

To summarize, this paper makes the following contributions:
\begin{enumerate}
    \item Techniques for transforming designer comments, sketches, and revisions into machine-learnable preference pairs.
    \item A dataset of 1460 UI screens associated with design feedback collected from twenty one designers. An analysis of this data shows UI preference pairs generated from natural designer feedback workflows have lower levels of disagreement than conventional ranking feedback.
    \item \textcolor{black}{A validation of our approach that shows how designer feedback improves open-source UI generation models and a quantification of the quality-quantity tradeoff for collecting designer feedback.} An arena-style evaluation showed that some forms of designer feedback, such as sketches and revisions, were highly effective in improving open-source models. Our best-performing model outperformed all tested baselines, including a larger proprietary model, GPT-5.
\end{enumerate}

\textcolor{black}{To promote reproducibility, we released the dataset and models from our work.\footnote{\url{https://github.com/apple/ml-rldf}}}

\section{Related Work}
To contextualize our work, we review literature in i) UI generation, ii) annotation interfaces for collecting human feedback, and iii) and tools to support designer workflows.
\subsection{UI Generation}
There have been numerous approaches developed to dynamically generate UI.
Model-based user interface development (MBUI) is a development approach for user interface applications, where developers first describe desired UIs in abstract specifications~\cite{traetteberg2002model,puerta1998towards}, which an MBUI environment then uses to generate concrete code implementations.
Mobi-D~\cite{puerta1997mobi} is an example of a MBUI system that separated UI development into different models corresponding to application data, user interactions, and presentation strategy.
The Personal Universal Controller is another example that was developed to generate personalized interfaces for different appliances, using a unified remote control device~\cite{nichols22}.

Artificial intelligence (AI) approaches have also been used to make dynamic, data-driven decisions on how this model to implementation should occur.
SUPPLE is a toolkit developed to dynamically personalize UIs based on a set of constraints and objective functions that represent device affordances~\cite{gajos2005fast,gajos2004supple}, user preferences~\cite{gajos2005preference}, and user ability~\cite{gajos2008improving}.
More recently, machine learning approaches, especially those using data driven neural networks, have been applied to code generation.
These large models~\cite{achiam2023gpt,roziere2023code,lozhkov2024starcoder} are promising in that they are trained on large amounts of data, which allows them to learn the distribution of text and code. %
Several works have finetuned LLMs for layout and UI-related tasks~\cite{tang2024layoutnuwa,10.1007/978-3-031-72661-3_10,10.1007/978-3-031-73007-8_26,wu2024uicoder}, often using custom datasets.
UICoder is an LLM that generates SwiftUI code using automated tools such as code compilers and vision-language models as training signals~\cite{wu2024uicoder}.
While this training approach was effective at increasing the syntactic understanding of LLMs, improvement was more difficult for design-related aspects, and the authors showed that models still made numerous design-related errors.
Overall, data-driven approaches to UI generation would benefit from greater volumes of human feedback, specifically from domain experts such as designers.

\subsection{Interfaces for Collecting Human Feedback}
One way that machine learning models have improved over time is by using human expertise to provide labeled data for training. Both domain experts and non-experts have used annotation interfaces to provide labels for model training. A body of work studies the design of annotation interfaces for humans to provide their knowledge to models. For example, the Teachable Machine is a system where people interactively provide a small number of demonstrations to train a machine learning model~\cite{carney2020teachable}.

Researchers have designed some of these interfaces to maximize the speed in which people can provide labels and enable them to collaborate to solve tasks~\cite{little2010turkit}. Other work has designed interfaces to make annotation more enjoyable, such as Games with a Purpose (GWAP), which let humans collaborate to solve real-world, computationally difficult tasks~\cite{von2006games}. In contrast to our goal for this paper, most of these works are not targeted towards enabling experts to provide feedback through their existing workflows. 

More recently, research has introduced interfaces to allow people to provide feedback to LLMs, which are unique in their breadth of supported tasks and expressive output space. Some of these interfaces allow people to rank LLMs output which is later used for training and evaluation. The LMSYS arena is an interface for crowdsourcing LLM evaluation by having people provide pairwise evaluations~\cite{chiang2024chatbot}.
The LMSYS arena has several variations for people to provide feedback for specific applications like design, code generation, and multi-modal conversations.
\textcolor{black}{Several interfaces have been developed to accelerate or improve these ranking processes, through visual annotations~\cite{gebreegziabher2024mocha},LLM response decomposition~\cite{shi2025dxhf}, clustering visualizations~\cite{kompatscher2025interactive}, and sensemaking approaches~\cite{gero2024sensemaking}.}

\textcolor{black}{Beyond ranking-style interfaces, other work has investigated collecting richer types of human feedback beyond binary signals such as natural language explanation, demonstrations~\cite{shaikh2025aligning}, and sequential design revisions~\cite{xie2024leveraging}.} %
\textcolor{black}{RLHF-Blender is an example of an interface that supports numerous types of feedback for RLHF-training of agents, including evaluative feedback, comparative feedback, corrective feedback, demonstrative feedback, and descriptive feedback~\cite{metz2023rlhfblender}.}
\textcolor{black}{In our work, we specifically investigate interfaces for collecting designers' feedback on UI designs, inspired by their everyday activities, such as commenting, sketching, and direct revision.}

\subsection{Tools for Designers}
Since our work focuses on types of natural feedback that can be captured from designers' workflows, we review computational tools built to support designers' workflows. 
In the book, Sketching User Experiences, Buxton highlights the need for tools that facilitate iterative UI design and development~\cite{buxton2010sketching}.
Several lines of research have focused on building tools that process low-fidelity designer artifacts, such as sketches and mockups, and transform them into higher-fidelity UIs with code implementations~\cite{landay1995interactive,huang2019swire,beltramelli2018pix2code}.
For example, SILK was a system that translated hand-drawn sketches of UIs into interactive, testable prototypes using gesture-recognition software~\cite{landay1996silk}. 
Swire uses a neural network encoder to retrieve relevant high-fidelity examples from a UI database for design inspiration~\cite{huang2019swire}.
This body of work suggests that artifacts already produced by designers (e.g., sketches) can encode information valuable for producing high-quality UIs.

Other research investigates how designers critique and revise UIs and provides tools to support this process.
d.note helped designers revise UIs through change tracking, annotations, and supporting design revisions~\cite{hartmann2010d}.
Compared to traditional workflows involving sketching on static images, d.note allowed designers to more efficiently implement suggested changes with fewer clarifications.
Charrette supported designers in giving feedback during design reviews and meetings~\cite{o2018charrette}.
Charrette classified the types of annotations that occur on \textit{artboards}, digital canvases used by designers to create, edit, and present designs, and the authors developed a web-based application that facilitates iteration and discussion of UI designs.
Insights from these tools suggest that in many cases, designer workflows can be computationally supported and encoded.

\section{Background} \label{sec:background}
Our paper focuses on a formulation of UI generation where an LLM model is provided with a textual prompt (e.g., a natural language description), and generates a UI represented as code (e.g., HTML).
Since we aim to improve their UI design capabilities, we first provide technical background on i) the data needed to train LLMs and ii) opportunities for designer input to improve this data.

\subsection{Data Format}
LLMs are typically trained in multiple stages and can incorporate different types of data in each~\cite{ouyang2022training}: i) unsupervised pretraining, ii) supervised finetuning, and iii) model alignment.

Most of an LLM's training time is spent during the \textit{pretraining} stage, where the model uses an unsupervised objective to learn the distribution of text and other data from large, unstructured data, such as web dumps.
During the \textit{supervised finetuning} stage, the model is trained to replicate a human-authored output given a distribution of input prompts, which allows the model to follow instructions rather than simply predicting continuations.
Finally, the \textit{model alignment} stage, focuses on fine-tuning models so that they more closely match human preference data.
Instead of training on a single ``ground truth'' response, as in the previous stage, alignment training relies on numerical ratings or rankings of output candidates, which is more useful for subjective tasks.

Of the two stages of LLM training where it is possible to incorporate human input, we choose to focus on the \textit{model alignment} stage, since designers tend not to directly write code implementations needed for constructing input/output examples and there might be numerous possible ``ground truth'' responses for a single prompt.
Therefore, \textbf{our paper focuses on collecting paired comparison data from designers to model their design preferences.}

\begin{equation}
\mathcal{D}_{\mathrm{pref}}
\;=\;
\{\, (x,\, y^{+},\, y^{-}) \;:\; y^{+} \succ y^{-} \}
\label{eq:preference-data}
\end{equation}

Formally, the dataset of design preferences $\mathcal{D}_{\mathrm{pref}}$ consists of triplets of an input natural language UI description $x$ and two possible UI designs $y^+$ and $y^-$, where $y^+$ (``chosen'' sample) is rated by a designer to be preferred over $y^-$ (``rejected'' sample).

\subsection{Data Quality}
Previous research has shown that, in practice, the quality of data has a significant impact on model performance~\cite{lambert2024t,xu2024dpo,10.5555/3666122.3668142}.
It may seem straightforward to generate paired comparison data entirely synthetically.
For example, a simple strategy would be to use a large, strong model (e.g., GPT-5) to generate chosen responses and a small, weak model (e.g., GPT-1) to generate rejected responses.
However, these ``trivial'' pairs may be less effective, since during training, they cannot help the model assess and choose between outputs with more subtle differences.

Due to designers' valuable expertise, there have been several attempts to include them in model training.
For example, previous work~\cite{duan2024generating} collected natural language critiques from designers who were asked to follow a pre-defined rubric.
However, the authors found that they disagreed with a significant portion of critiques generated by models trained on this data and even other designers.
Other work~\cite{wu2024uiclip} employed a similar approach where designers were asked to follow a high-level rubric and provided binary comparison ratings of sampled UI pairs.
However, this also led to low inter-rater reliability and significant amounts of label noise, which negatively impacted model performance.
We hypothesize that a key reason is a mismatch between the subjective nature of UI design and the rigidity of rubric-based rating tasks.
When evaluating synthetic outputs for model alignment, designers often must choose between two flawed options, each with different strengths and weaknesses.
This process reduces their nuanced expertise to coarse labels and offers little opportunity to propose concrete improvements.
Instead, in this paper, \textbf{we show that collecting data from designers models around their existing workflows leads to higher quality data and better model performance.}

\section{Methodology}
\begin{figure*}[!htb]
    \centering
    \includegraphics[width=\linewidth]{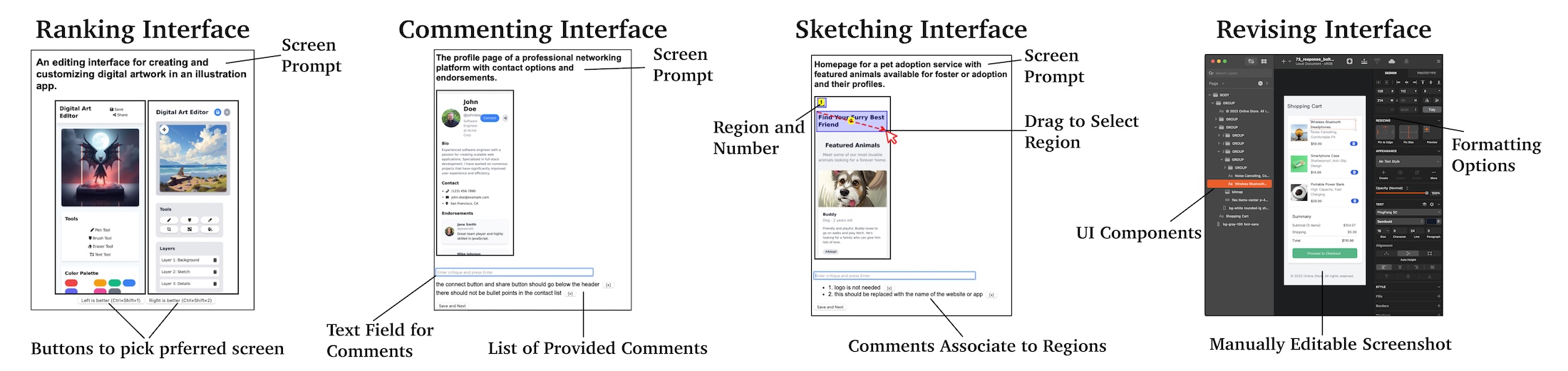}
    \Description{Four screenshots of our four labeling interfaces for ranking, commenting, sketching, and revision. Ranking displays two screenshots with two buttons to choose the better screen. Commenting shows a single screenshot with a textbox for adding to a list of design comments. Sketching shows a single screenshot where the user is dragging a box to capture a region of the screenshot. Revision shows a screenshot of a user editing a UI design in sketch.}
    \caption{Figure shows the four interfaces we developed to collect feedback from designers. \textmd{The ranking interface (Far Left) allows users to select the better of two UI screenshots through a binary response. The commenting interface (Center Left) allows users to write a list of natural language critiques or comments for a UI screenshot. The sketch interface (Center Right) allows users to draw annotations (boxes and points) on a UI screenshot and associate them with textual comments. Designers used the Sketch design software (Far Right) to make direct edits to model-generated UIs, which were first converted into the appropriate format. The commenting, sketching, and revising interfaces and inspired by interactions identified by Hartmann et al~\cite{hartmann2010d}.}}
    \label{fig:feedback-interactions}
\end{figure*}
We describe our approach for i) generating UI code, ii) building interfaces to collect designer-aligned feedback for those UIs, and iii) using the feedback to finetune UI code generation models.

\subsection{Initial Data Generation}\label{sec:initial_data_generation}
We first generated a large UI dataset to provide materials for designers to annotate using on-policy sampling of a base model.
At a high-level, our approach consisted of i) generating a list of textual descriptions of UIs, ii) using a code LLM to generate UI programs for the textual descriptions, and iii) rendering the UI programs into screenshots.

Instead of using existing datasets of UI descriptions~\cite{wang2021screen2words}, we synthetically generated a list of diverse and detailed descriptions to seed our UI generation. %
We prompted an LLM to generate a large list of approximately 100,000 natural language descriptions.
To generate this list, we first prompted the LLM with a set of manually-authored example descriptions that described UI functionality, layout, and content, and then asked the LLM to generate 10 more unique descriptions.
We evaluated this prompt using a high temperature value, which increased the probability of the LLM generating a new set of descriptions.
We merged descriptions from each evaluation into a global set until the number of descriptions reached our target.

To generate UI programs corresponding to these prompts, we used a publicly released code LLM, Qwen2.5-Coder 32B.
At the time we started this project, Qwen2.5-Coder was the strongest publicly available code LLM that could fit on a single GPU.
We sampled a random description from the list of UI descriptions and prompted the code LLM to generate a well-designed web page using a limited set of web libraries (HTML, Tailwind, and Font Awesome).
We used each randomly-selected description to generate 32 different possible web pages via temperature-based sampling of the LLM.

We rendered each UI program into a screenshot using an automated pipeline.
First, a HTML parser extracts all referenced images in the code (i.e., \texttt{<img>} tags). To generate plausible placeholder assets for rendering images, we fed each image tag's \texttt{alt} attribute into an off-the-shelf text-to-image model called Flux Schnell. The HuggingFace ID for this model is \texttt{black-forest-labs/FLUX.1-schnell}. 
Our code generation prompt contained explicit instructions to include alt-text for all images; however, if the \texttt{alt} tag was still missing, we used the \texttt{src} attribute value instead.
We developed an automated script that i) staged each UI's HTML code and required assets (e.g., library files and image assets) into a server then ii) used a headless browser to visit the hosted URL and take a screenshot.
In addition to the browser-based screenshot rendering, we developed a program based on the open-source \texttt{html2sketch} library to convert each web page into an editable Sketch file using the computed positions and styles from the browser DOM.
To further improve this dataset, we used the UIClip~\cite{wu2024uiclip} base model to compute the quality scores of UIs.
For the 32 original UIs generated for each description, we kept the top 8 outputs according to their computed quality score.
This allowed many ``obviously bad'' outputs to be filtered out automatically, which allows designers to give more subtle feedback on plausible screens.

To summarize, we generated a dataset of \textasciitilde6400 UIs corresponding to \textasciitilde200 unique textual descriptions.
Each UI contains a natural language description, HTML source code, a screenshot rendered by a browser engine, and a Sketch file.

\subsection{Designer Feedback Interfaces}

To collect feedback from designers on the generated UIs, we designed four annotation interfaces inspired by the design principles identified by Hartmann et al.~\cite{hartmann2010d} (Figure \ref{fig:feedback-interactions}).
Hartmann et al. conducted a broad literature review of how designers revise artifacts in textual documents, source code, movies, games, and user interfaces.
Based on their review of these domains, they proposed four principles important for UI revision, of which three correspond to actions by designers themselves: i) commenting, ii) sketching, and iii) revising. We developed annotation interfaces for i-iii, and included a baseline ranking interface inspired by current practices for incorporating user feedback into LLMs through pairwise comparisons. 

\subsubsection{Ranking}
We designed the baseline ranking interface to mimic strategies used by LLM chat interfaces to elicit user feedback, e.g., asking users to select the better of two generated responses.
Some other common interactions used to collect rankings include "thumbs up" or "thumbs down" controls, and implicitly ranking responses through user-initiated response regenerations (e.g., user is not satisfied with the first response).
Our ranking interface displays a textual description and two candidate UIs arranged side-by-side.
\textcolor{black}{The two candidates are selected using uniform random sampling without replacement, following practices used in LLM evaluation and ``judge'' calibration settings~\cite{chiang2024chatbot,zheng2023chatbotarena}.}
The interface (Figure \ref{fig:feedback-interactions} Far Left) asks designers to select the UI that they feel is better-designed.
If both UIs are poorly designed (e.g., both contain many design flaws), the interface asks designers to choose the UI that they feel is a better starting point for fixing.

\subsubsection{Commenting}
Multiple studies suggest that designers often write high-level comments as critiques, propose changes, or ``todo items~\cite{hartmann2010d,o2018charrette}.'' %
While designers use a wide variety of tools (e.g., text editors, note-taking software), for the purposes of our experiment, we developed a commenting interface that allows designers to provide natural language feedback on UIs.
The interface (Figure \ref{fig:feedback-interactions} Center Left) presents designers with a UI screenshot and its textual description, and asks them to write a list of natural language critiques.
Designers type each critique or suggestion into a text-field then hit the Enter key to add it to a list.

\subsubsection{Sketching}
The sketching interface (Figure \ref{fig:feedback-interactions} Center Right) is similar to the commenting interface; however, the sketching interface allows designers to provide visually grounded feedback (e.g., annotations).
The interface presents designers with a UI screenshot and asks them to identify areas for improvement within a screenshot.
The interface allows drawing bounding box and point annotations on top of the UI.
After drawing an annotation, the interface asks the designer to provide a natural language feedback for that region.
Designers can draw multiple annotations on the UI screenshot which it displays as a list on the interface.
\subsubsection{Revising}
In addition to providing critiques or feedback, designers often use direct manipulation to revise UI designs in software such as Figma and Sketch.
Previous work suggests that in many cases, designers may prefer this revision to occur in the output domain (e.g., rendered graphics) rather than the source domain (e.g., source code)~\cite{hartmann2010d}. %

We developed an interface that allows designers to revise LLM-generated UIs using Sketch (Figure \ref{fig:feedback-interactions} Far Right).
The interface first displays a UI's description and screenshot to allow designers to think about possible flaws and potential fixes.
The interface asks designers to download the corresponding Sketch file, which we generated from our initial data generation process.
Designers modified the Sketch file to improve the UI, and then uploaded it back to the interface.

\subsection{Model Training}
We developed a pipeline to convert designer feedback data into preference pairs and used them to improve UI generation models through fine-tuning.

\subsubsection{Data Preprocessing}
Because designer feedback is often not in the format required for model alignment (Section \ref{sec:background}, we used several strategies to convert them into preference pairs.
\begin{itemize}
    \item \textbf{Ranking.} For our baseline, we directly used designers' rankings to form preference pairs out of the UI screenshots.
    \item \textbf{Commenting.} Designers produced a list of natural language comments corresponding to a UI screenshot. For each screenshot, we prompted an LLM to improve the UI's HTML code using the list of provided comments. We then re-rendered the code into a new screenshot, which we labeled as preferred over the original UI screenshot. \textcolor{black}{The model and prompt used for generating the improved UIs can be found in the appendix of this paper.}
    \item \textbf{Sketching.} Designers produced a list of visually-grounded annotations corresponding to a UI screenshot. We adopted a similar approach to processing commenting data, however we associated each textual comment with an HTML code snippet of the DOM element with the highest overlapping IoU score with the drawn box. \textcolor{black}{The model and prompt used for generating the improved UIs can also be found in the appendix.}
    \item \textbf{Revising.} Designers downloaded a Sketch file generated from the code used to render a UI screenshot then modified and re-uploaded an improved Sketch file. We formed a preference pair using the rendered preview of the improved Sketch file as the preferred sample and a rendered preview of the original Sketch file as the alternative.
\end{itemize}
\subsubsection{Training Pipeline}
Our training setup was similar to existing RLHF architectures~\cite{ouyang2022training} that involve training two models: i) a reward model and ii) a generator model.

Our setup used a two-step process: we first trained the reward model using human feedback, then we used this reward model as a training signal for training the generator.
Previous work~\cite{ouyang2022training} found that this two-step approaches provides advantages over directly using human feedback to train the generator, such as improved sample efficiency.
Reward models are typically trained to assign a numerical score to an input/output pair from the generator (e.g., the concatenation of a textual prompt and response).
The reward model was trained to assign higher scores to ``preferred'' responses marked by human labels over ones that are not.
After training, the reward model was used to guide the generator model to produce outputs that maximize its score, either through traditional reinforcement learning or data generation. %
More details can be found in other papers~\cite{ouyang2022training,lambert2024t}.

\paragraph*{Reward Model.}
We chose to base our reward model on the multi-modal \texttt{CLIP B/32} architecture, which we initialized from the publicly released UIClip model~\cite{wu2024uiclip} to improve training efficiency.
To isolate the effectiveness of our training data, we used the UIClip checkpoint which was not trained with any human preference pairs from their original paper.\footnote{HuggingFace ID of our initialization checkpoint \texttt{biglab/uiclip\_jitteredwebsites-2-224-paraphrased}}
The reward model accepts i) a rendered image (a UI screenshot) and ii) a natural language description (a target description of the UI).
These two inputs are fed into the model to produce a numerical score (reward), which is calibrated so that better-quality visual designs result in larger scores.
To assign rewards to HTML code, we used the automated rendering pipeline described in Section \ref{sec:initial_data_generation} to first render code into screenshots using browser automation software.

To train the reward models using our designer feedback dataset, we used a margin-based variation of the original pairwise contrastive objective in UIClip, which tunes the model so that it assigns higher scores to ``preferred'' samples over ``rejected'' samples.
\begin{equation} \label{eq:margin_loss}
\mathcal{L}(s^+,s^-)=\max\left\{0,\ s^--s^++m\right\}
\end{equation}
In our loss function (Equation \ref{eq:margin_loss}), $s^+$ and $s^-$ refer to the reward model scores of the preferred and non-preferred UI screenshots, respectively. These scores were computed by computing a scaled dot product between the encoded textual description and UI screenshots.
$m$ refers to an empirically determined margin value.

When finetuning, we froze all but the last layer to prevent overfitting and employed an additional data augmentation technique.
Because we only asked designers to give feedback on the top 8 out of 32 outputs (Section \ref{sec:initial_data_generation}), it caused our dataset to exclude many examples of poor UIs, which are nevertheless needed to accurately represent the entire distribution of generated UIs.
To resolve this imbalance, we sampled additional UI pairs from the entire output distribution (i.e., from all 32 candidate outputs) and synthetically labeled them using UIClip's score.
The reward model was trained using both designer-labeled preference pairs and these UIClip-labeled pairs, which were sampled with a fixed probability.

All reward models used the same hyperparameters and were trained for a fixed number of optimization steps.
We determined the hyperparameters for the reward model training by manual inspection and experimentation, and they are provided in the appendix of this paper.

\paragraph*{Generator Model.}
We finetuned generator models to generate outputs that maximize the expected reward using an optimization algorithm called ORPO~\cite{hong2024orpo}.
The ORPO algorithm expects a dataset of triplets consisting of an input prompt (textual description) and two candidate outputs (HTML programs), where one of them is preferred over the other.
Note that this is similar to the format of our design preference dataset (Equation \ref{eq:preference-data}); however, ORPO's training data requires the output candidates to be code (e.g., HTML), not the UI images collected from designers. 
To produce the required data, we first used the generator model to synthesize a large batch of UI programs (HTML) from a list of descriptions, sampling multiple candidates (32) per description.
In total, we generated \textasciitilde400{,}000 HTML UIs from \textasciitilde11{,}600 randomly sampled unique textual descriptions.
Following this code generation, these HTML programs were assigned numerical scores by first rendering them into screenshots then running the reward model over them.
With this dataset of scored HTML programs, we followed a procedure from previous work~\cite{tunstall2023zephyr} to generate preference pairs by using the top response for an input description as the ``chosen'' response and a randomly selected HTML program from the same description as the ``rejected'' response. %
Finally, we ran the ORPO algorithm~\cite{hong2024orpo} for one epoch to learn on preference pairs labeled by the reward model.
When finetuning, we employed parameter offloading~\cite{ren2021zero} and mixed-precision training to improve efficiency.

All hyperparameters for the generator model training were determined by manual inspection and experimentation, and they are provided in the appendix of this paper.

\section{Evaluation}
We conducted three evaluations to investigate i) the quantity and quality of data collected from designer feedback interfaces, ii) the relative performance of models trained using different forms of designer feedback, and iii) the generalizability of designer feedback for model fine-tuning.
\subsection{Designer Feedback Evaluation}

To obtain examples of designer feedback using our feedback interfaces, we recruited twenty one designers at our institution, a large technology company.
Our design feedback evaluation investigated the quality and quality of data collected from these interfaces when converted to preference pairs.

\subsubsection{Participants}
We recruited participants by posting messages in company message boards and through word-of-mouth.
The recruited participants had varying levels of professional design experience, ranging from 2 to over 30 years.
Participants also worked in different areas of design, such as UI/UX design, product design, and service design.
Participating designers also noted the frequency of conducting design reviews (both formal and informal) in job activities: ranging from once every few months to multiple times a week.

\subsubsection{Procedure}
During the study, participants joined a video call where a member of the research team first gave an overview of the study and asked for their informed consent.
Participants used each of the four feedback interfaces in randomized order, where they first watched a pre-recorded tutorial video (1-2 minutes) for each feedback interface then spent 10.5 minutes giving feedback to UIs using that feedback interface. In total, the session lasted approximately one hour.
As a thank-you for participating, we offered participants a meal voucher.

\subsubsection{Post-study Interview}
Since the primary goal of this work is to develop designer-aligned interactions for providing feedback to ML models, we first validated the set of designer interactions we chose from previous work~\cite{hartmann2010d}

In our post-study interview, we asked participants to estimate the amount of time that they spent doing similar tasks.
Overall, participants estimated that they spent the most time on revision-like tasks (average of 33\%), and participants estimated they spent the least time on ranking-like tasks (9\%).
Commenting (26\%) and sketching (26\%) also constituted a significant amount of participants' design activities, validating that our chosen interaction strategies are representative of the types of tasks that designers normally do.
Since designers report doing other types of activities in their jobs, these percentages do not add up to 100.
Most participants who were primarily UI/UX designers felt that editing and revising designs (e.g., direct manipulation in design editing software), took a significant amount of their time, and sometimes referred to these activities as ``hands-on design work.''

\subsubsection{Dataset Overview}
In total, we collected 1460 annotations from twenty one designers, where one ``annotation'' refers to a single UI screenshot paired with all the feedback for that screen, \textit{e.g.,} multiple comments.
One annotation maps to one preference pair for model training.

Designers collected the most annotations using the ranking interface (1063), and designers, on average, were able to generate 4.8 rankings per minute.
Designers collected the least annotations using the revision interface (64), on average requiring 3.45 minutes per revision.
We expected this discrepancy in samples, since this condition required participants to attempt to fix UIs rather than just evaluate them.

The sketching interface (181) led to more annotations than the commenting interface (152).
We hypothesized this is because the commenting interface required designers to type longer text (87.1 characters on average) than the sketching interface (42.2 characters on average), since the designers needed to textually describe UI elements instead of being able to annotate the element directly on the image. 
Furthermore, the sketching interface generally led to more feedback (\textit{e.g.,} number of comments) per UI (2.7 on average) than the commenting interface (1.9 on average).
Examining the annotations and comments, we found that comments from the sketching interface typically contained lower-level feedback about particular regions (\textit{e.g.,} ``make this text larger'') while the commenting interface typically contained higher-level feedback (\textit{e.g.,} ``the screen has poor information hierarchy'').

\begin{shaded}
The annotation interface designers used strongly impacted the number of data samples that could be collected in a fixed duration. On average, designers collected over 15x more annotations using the ranking interface than the revision interface.
\end{shaded}

\subsubsection{UI Preference Pair Quality Assessment} \label{sec:data_noise_analysis}
We conducted a quality assessment with six members of the research team (who are HCI experts) to estimate the quality of UI preference pairs generated from each condition of the designer feedback evaluation.
Since there is no standardized rubric or rating system to evaluate the quality of UI preference pairs, we operationalize quality as the percent agreement between the researchers' choice of the best screen in each UI preference pair, and the ``improved'' UI from the UI preference pairs generated from designer feedback. 

In other words, if both an HCI expert and a design professional independently agree on the best screen for a UI preference pair, that UI preference pair ranking is more likely to be accurate. 

During this evaluation, we sampled UI preference pairs from the designer-generated dataset, and stratified them by the feedback interface used to collect them.
In a web interface, we showed example pairs to a researcher who did not have knowledge of i) the type of feedback used to generate the pair and ii) which screenshot in the pair was ``preferred'' by the designer.
The interface told researchers to choose the UI from each pair that they felt was better designed, and we measured the rate at which the researcher's choice of the best screen corresponded to either i) the UI preferred by designers during the ranking task, ii) the improved UI from the designer's edits during the revision task, or iii) the improved UI generated by running an LLM on the designer's comments or sketch annotations from the commenting and sketching task, respectively.
Each researcher spent around 30 minutes on this task, which led to 695 ratings in total.
\subsubsection{Results}
Overall, researchers agreed with designers for 61.7\% of the UI preference pairs. 
Of the four feedback conditions, researchers most often agreed with the designer on the best screen from pairs generated from the revision condition (76.1\%), which suggests that allowing designers more control reduces label noise.
On the other hand, researchers and designers had very low agreement on the best screen (49.2\%, close to random chance) from pairs generated from the ranking condition. 
The low agreement for ranking-style data is consistent with prior work that compared designer pairwise rankings and also found low agreement among independent designers~\cite{wu2024uiclip}. %
Compared to prior work, we expected lower agreements from our ranking experiment because the pairs used were already pre-filtered to reduce ``easy'' comparisons with obvious design flaws (Section \ref{sec:initial_data_generation}).

In general, designers and researchers had higher agreement on their choice of the best screen from preference pairs generated from the comment and sketch conditions, likely because they provided opportunities for designers to give specific feedback on flaws and how to fix them.
However, the quality of these UI preference pairs may have been limited by the LLM's ability to operationalize some types of high-level designer comments, \textit{e.g.,} ``improve the information hierarchy of this screen''. 
It is reasonable to expect that future, stronger LLMs will possess better code editing abilities which would in turn improve the quality of comment-derived data.
Researchers and designers agreed more often on the UI preference pairs generated from the sketching condition (63.6\%) than the UI preference pairs generated from the commenting condition (57.3\%). We hypothesized this may be because including grounded annotations for each comment on a UI design enabled the LLM to better localize the area of the UI the comment applied to, and thus generate a better fix for it.

\begin{shaded}
Researchers and designers had varied levels of agreement on UI preference pairs generated as output from designer feedback on different annotation interfaces. We use agreement as a proxy for data quality. \textit{For our chosen model}, UI preference pairs generated from designers' revisions led to the highest agreement while UI preference pairs generated from designers' pairwise rankings led to the lowest agreement.
\end{shaded}

\subsection{Feedback Fine-tuning Evaluation} \label{sec:feedback_comparison_study}
Next, we conducted an evaluation to empirically determine the most effective source of designer feedback for model training.
While our data quality assessment showed that data collected from designers' revisions of UIs led to the highest agreement rate, other forms of feedback (e.g., sketching), could be collected significantly faster, and the additional data volume could potentially offset the slightly higher noise when training models. \textcolor{black}{The purpose of this evaluation is to investigate this tradeoff.}
Using our training approach, we trained several versions of code generation models from different types of designer feedback and two baseline conditions.

In total, we evaluated 6 conditions:

\begin{itemize}
    \item Qwen2.5-Coder - The 32B variant of the Qwen2.5-Coder model,\footnote{Ollama ID \texttt{qwen2.5-coder:32b-instruct-fp16}}. This was the base generation model that generated UIs that were shown to designers. This baselines represents model performance without any additional UI-specific fine-tuning.
    \item Qwen2.5-Coder + UIClip - Qwen2.5-Coder trained using the base model of UIClip~\cite{wu2024uiclip} as a reward model. This baseline has undergone additional UI-specific fine-tuning with an off-the-shelf reward model trained without any designer feedback.
    \item Qwen2.5-Coder + Ranking - Qwen2.5-Coder fine-tuned using a reward model derived from ranking data.
    \item Qwen2.5-Coder + Comment - Qwen2.5-Coder fine-tuned using a reward model derived from comment data.
    \item Qwen2.5-Coder + Sketch - Qwen2.5-Coder fine-tuned using a reward model derived from sketch data.
    \item Qwen2.5-Coder + Revision - Qwen2.5-Coder fine-tuned using a reward model derived from revision data.
\end{itemize}

Samples from these models can be found in the Appendix, Section \ref{sec:example_outputs}.

\subsubsection{Procedure}

To evaluate the performance of models trained under different conditions (e.g., types of designer feedback), we employed an arena-style evaluation with human judges, which has been used for general~\cite{chiang2024chatbot} and UI-specific~\cite{wu2024uicoder,webdev_arena} LLM model evaluation.
An arena-style evaluation compares the relative performance of several models by repeatedly sampling a pair of the pool of models, using both models to generate output for the same input prompt, then asking a human judge to select a ``winner'' from that pair.
Following existing practices, we computed rating scores from judges' responses.
These scores, also referred to as Elo ratings, are numerical estimations of each model's performance against other models in the same arena~\cite{chiang2024chatbot}.
We use the same approach and parameters set by LMSYS Chatbot ~\cite{chiang2024chatbot}, using their publicly available code~\cite{zheng2023chatbotarena}.
This approach scales Elo ratings to a standardized range, centered roughly at 1000, and calculates confidence intervals using bootstrap sampling. %

We generated 210 descriptions for the evaluation, to align the size of our test set with those used by previous work~\cite{wu2024uicoder,wu2024uiclip} and other coding-related benchmarks~\cite{chen2021evaluating}. 
The list of descriptions was generated by prompting an LLM with one-sentence app screen descriptions (methodology described in Section \ref{sec:initial_data_generation}) and are available in supplemental material.%
We examined the list of descriptions used for evaluation to remove duplicates and ensure that they do not appear in any of the training data.
Although we did not detect any exact matches in the remaining data, it is possible that some descriptions in the two splits are semantically similar, \textit{e.g.,} ``a login screen for a banking app'' and ``a sign-in page for a financial planner.''
We consider these types of similarities acceptable, since both login screens and finance related applications are common types of UIs.

We ran all models in the arena using their default sampling parameters.
We converted models with downloadable weights to GGUF format, quantized them to 16-bit precision, and loaded them into Ollama\footnote{https://ollama.com/}, an open-source utility for managing and running LLMs.
For efficiency, we limited the maximum output length for all models to 4096 tokens, which covers the portion of code responsible for rendering the initial screen viewport seen and rated by annotators.
We generated screenshots from the model code output using the pipeline described in Section \ref{sec:initial_data_generation}.

\subsubsection{Participants}
To evaluate the models, six members of our research team (who are HCI experts) participated in the model evaluation as human judges.
We showed the judges a randomly selected textual description and the generated UI screenshot output of two randomly selected models, and we instructed the judges to select the UI that they felt had a better design.
Note that the evaluation interface shown to judges does not show which samples were generated by which models (\textit{i.e.,} a ``blind'' rating task), reducing the risk of bias towards any specific condition.

In total, judges conducted 405 pairwise comparisons between the UI output screenshots of two models.
Since we tested 8 different models, this led to approximately 27 comparisons for each of the $\binom{6}{2}=15$ possible model comparisons.

\subsubsection{Results}

\begin{figure}[!h]
    \centering
    \begin{subfigure}{0.48\textwidth}
        \centering
        \includegraphics[width=20pc]{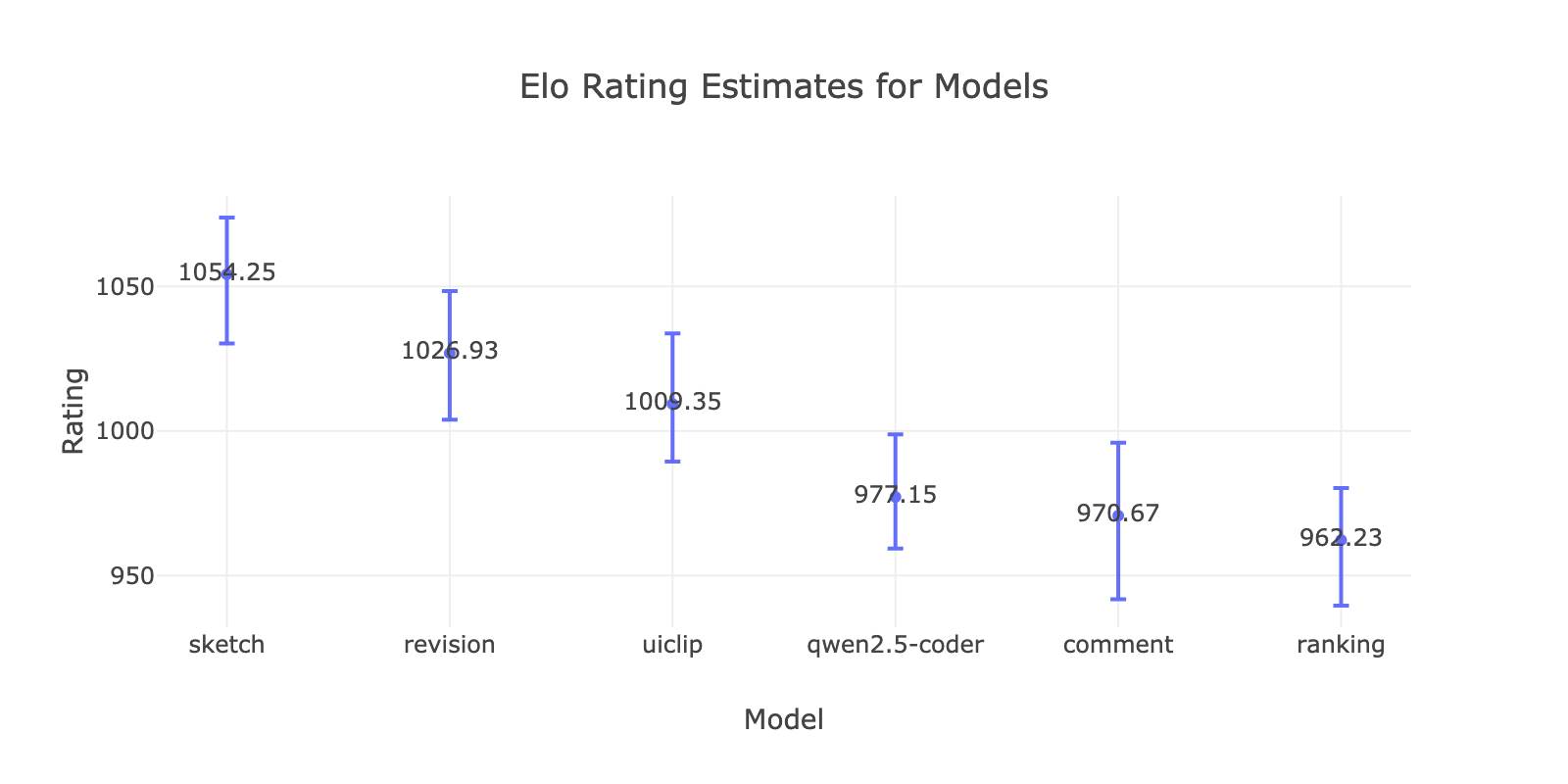}
        \Description{A confidence interval plot that shows each model's Elo ratings. The models from best to worst: sketch, revision, uiclip, qwen2.5-coder, comment, and ranking. Their ratings are 1054, 1026, 1009, 977, 970, 962, respectively.}
    \end{subfigure}
    \hfill
    \begin{subfigure}{0.48\textwidth}
        \centering
        \includegraphics[width=20pc]{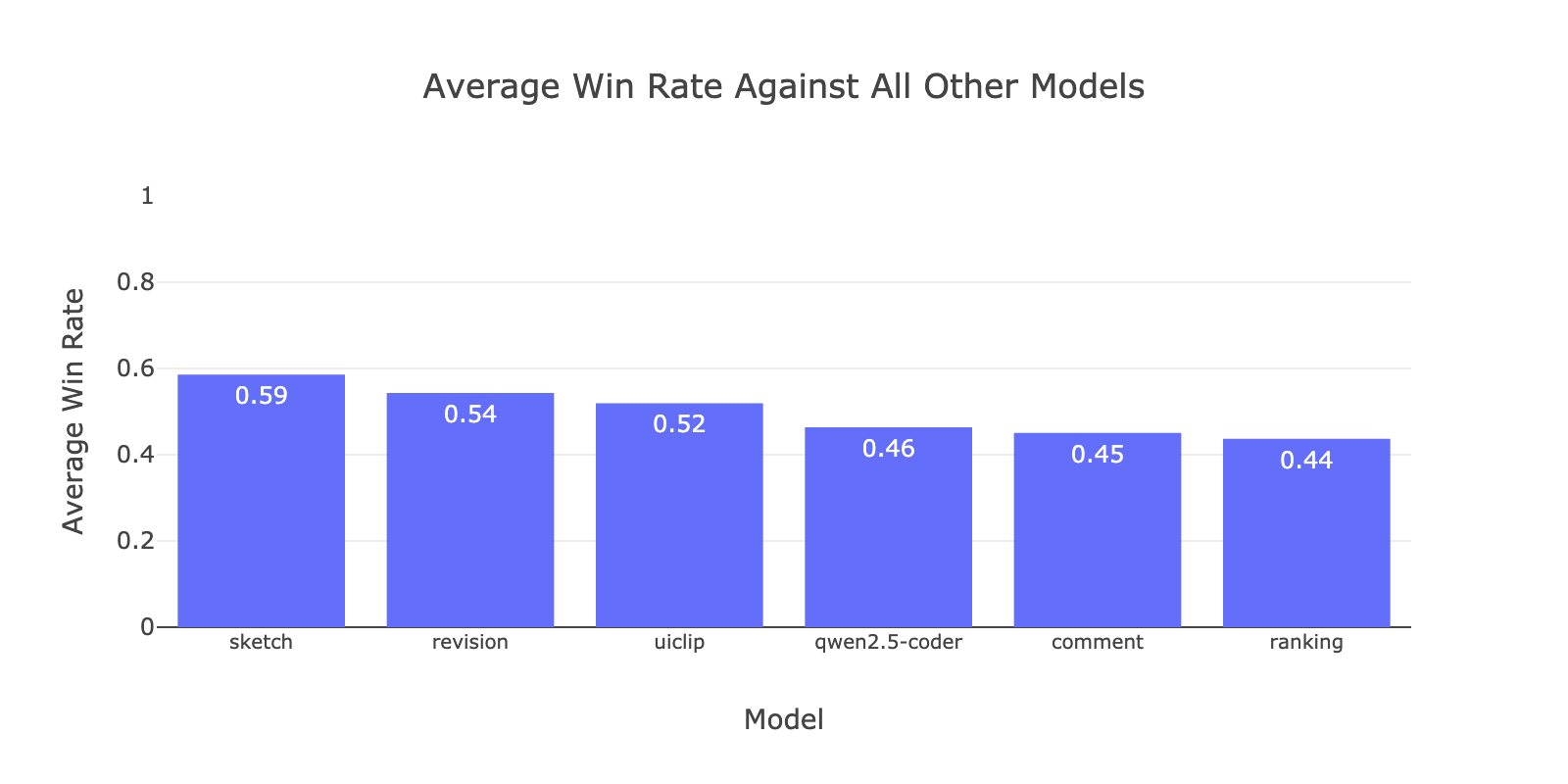}
        \Description{A bar chart that shows each model's average win rate against other models. The models from best to worst: sketch, revision, uiclip, qwen2.5-coder, comment, and ranking. Their win rates are 0.59, 0.54, 0.52, 0.46, 0.45, and 0.44, respectively.}
    \end{subfigure}

    \caption{Rating scores (Top) and average win rate (Bottom) of models in our feedback fine-tuning evaluation. 
    \textmd{We computed the rating scores by using the LMSYS calculation methodology~\cite{chiang2024chatbot,zheng2023chatbotarena}, and higher scores indicate models that were more often preferred by human judges. Bars show the median score and 95\% confidence intervals generated using bootstrap sampling.}}
    \label{fig:eloscores}
\end{figure}

Figure \ref{fig:eloscores} shows the results of our evaluation, which consists of the rating scores and confidence intervals for each model using the LMSYS calculation methodology~\cite{zheng2023chatbotarena,chiang2024chatbot}.
\textcolor{black}{The appendix of this paper (Section \ref{sec:wald_test}) contains additional statistical analysis, \textit{i.e.,} the p-values obtained through a Wald statistical test~\cite{wald1943tests}, which align with the visually observed confidence intervals.}

Overall, our results show that our fine-tuning approach can improve a code generation model's UI generation ability.
First, even training with an off-the-shelf reward model trained without designer feedback~\cite{wu2024uiclip} (i.e., UIClip condition) led to improvements over the un-tuned Qwen2.5-Coder model.
Because some pairwise comparisons were decided on obvious design flaws (e.g., overlapping text), UIClip’s training on synthetically introduced artifacts helped the generator avoid these errors.
We hypothesized that incorporating designer feedback would enable the model to better capture subtle design choices and assess trade-offs.
While some types of designer feedback were effective at further improving performance, others were not.

Among the tested designer-aligned conditions, the sketch and revision models performed better than the original base model, and the sketch model performed the best overall.
These scores roughly coincide with our result from our previous data quality assessment (Section~\ref{sec:data_noise_analysis}) which showed these types of feedback resulted in the highest agreement rates between researchers and designers (63.6\% and 76.1\% , respectively).
\textcolor{black}{The sketch model had a slightly higher rating than the revision model, although this was not statistically significant}.
We hypothesize that the sketch-trained model’s higher score over the revision-trained model stems from the roughly three times larger sketch dataset, which offset the revision data’s quality advantage.
In contrast, training on commenting or ranking data led to no change or even slight degradation in performance (Figure~\ref{fig:eloscores}), \textcolor{black}{although these slightly dips were not statistically significant.}
We hypothesize that the noisier labels and lower agreement from these feedback conditions led to a weak or detrimental signal for model training.

\begin{shaded}
The performance of UI generation models depends on both the quality and quantity of data used to train them. While some interactions require designers to spend more time and effort, they can also result in better data quality. \textit{For our chosen base model}, we found that data collected from sketching and revision feedback led to model improvements, and training on sketching feedback led to the best model performance.
\end{shaded}

\subsection{Model Generalization Evaluation}

Our feedback fine-tuning evaluation identified effective forms of designer feedback for model training; however, the prior experiment fine-tuned only the model that produced the designer-annotated data.
Because different base models may generate UI code with distinct distributions of design flaws (in both type and frequency), we conducted an evaluation to assess whether feedback derived from Qwen2.5-Coder 32B generalizes beyond that source model.
Therefore, we i) evaluateed generalizability by fine-tuning two additional open-source LLMs and ii) compared their performance against their untuned baselines and proprietary LLMs (e.g., GPT-5).
We selected Qwen2.5-Coder 3B~\cite{hui2024qwen2} and Qwen3-Coder~\cite{yang2025qwen3} to test generalization, since they were the best-performing permissively-licensed coding models capable of running locally on an edge device and consumer GPU at the time of this experiment, respectively.
From our visual inspection, their outputs qualitatively differ significantly from Qwen2.5-Coder 32B; however, because both are from the Qwen family, we acknowledge that this limitation potentially overstates our method's generalizability.

To train these base code generation models, we chose to use the best-performing sketch-trained reward model, which led to the best performance in our previous experiment.
It is possible that other forms of feedback may also be effective or exhibit different generalization properties, but we exclude them to prevent a combinatorial explosion of model variants.

In total, we evaluated the following 6 conditions:
\begin{itemize}
    \item Qwen2.5-Coder + Sketch - The Qwen2.5-Coder 32B model fine-tuned using a reward model derived from sketch data. This model performed the best in the previous feedback comparison arena, and we include it again as a point of reference.
    \item Qwen2.5-Coder 3B - The 3B variant of the Qwen2.5-Coder model.\footnote{Ollama ID \texttt{qwen2.5-coder:3b-instruct-fp16}}
    \item Qwen2.5-Coder 3B + Sketch - The Qwen2.5-Coder 3B model fine-tuned using a reward model derived from sketch data.
    \item Qwen3-Coder - The 30B variant of the Qwen3-Coder model.\footnote{Ollama ID \texttt{qwen3-coder:30b-a3b-fp16}}
    \item Qwen3-Coder + Sketch - The Qwen3-Coder model fine-tuned using a reward model derived from sketch data.
    \item GPT-5 - A multi-modal, reasoning-capable foundation model released by OpenAI~\cite{openai_gpt5_system_card_2025}. We set reasoning effort to ``medium,'' which is the default value in the API.
\end{itemize}
Samples from these models can be found in the Appendix, Section \ref{sec:example_outputs}.
For this experiment, we used the same evaluation procedure and participants as the previous feedback fine-tuning evaluation (Section \ref{sec:feedback_comparison_study}).

\subsubsection{Results}
\begin{figure}[!h]
    \centering
    \begin{subfigure}{0.48\textwidth}
        \centering
        \includegraphics[width=20pc]{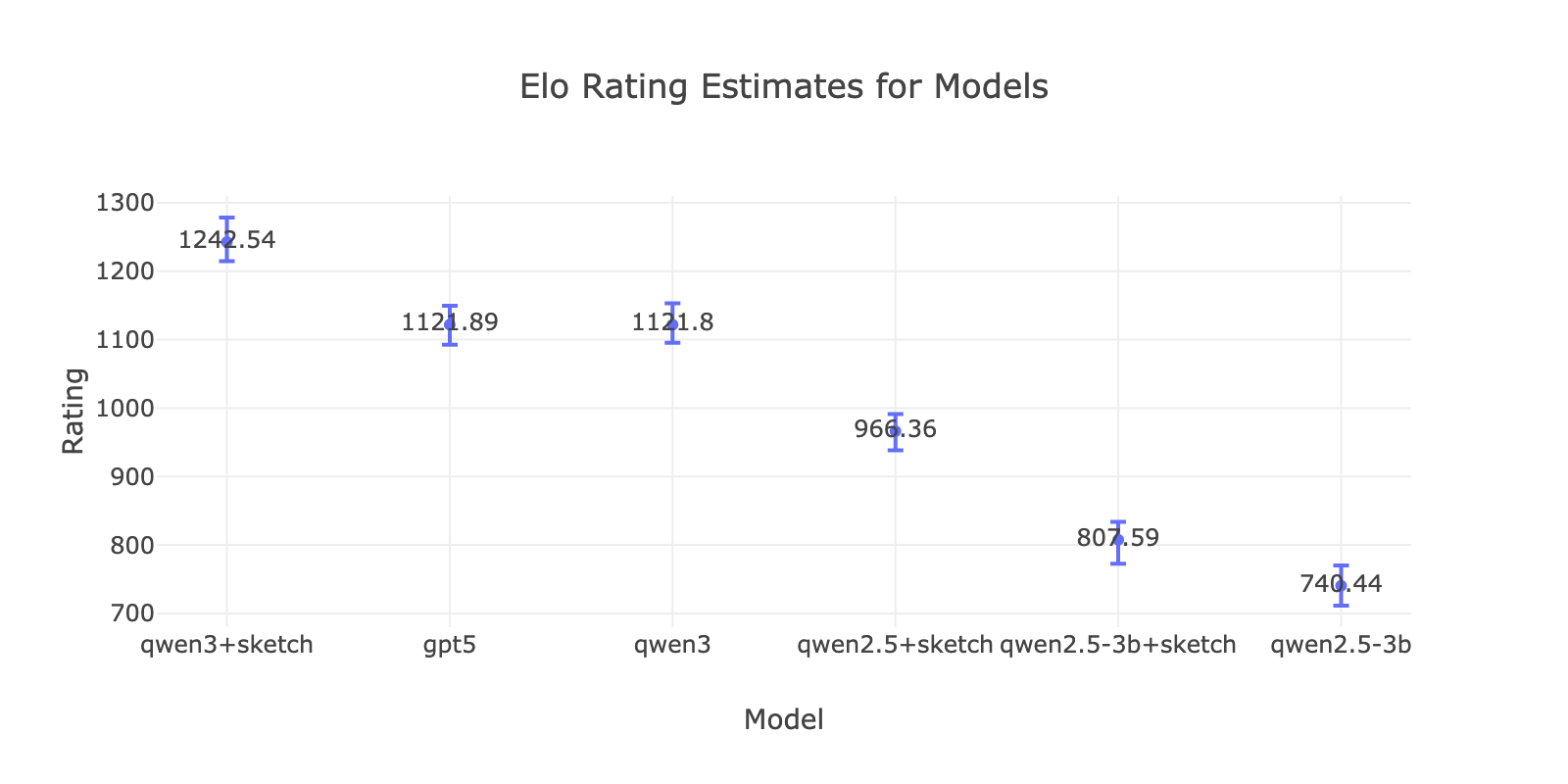}
        \Description{A confidence interval plot that shows each model's Elo ratings. The models from best to worst: qwen3+sketch, gpt5, qwen3, qwen2.5+sketch, qwen2.5-3b+sketch, qwen2.5-3b. Their ratings are 1242, 1121, 1121, 966, 807, 740, respectively.}
    \end{subfigure}
    \hfill
    \begin{subfigure}{0.48\textwidth}
        \centering
        \includegraphics[width=20pc]{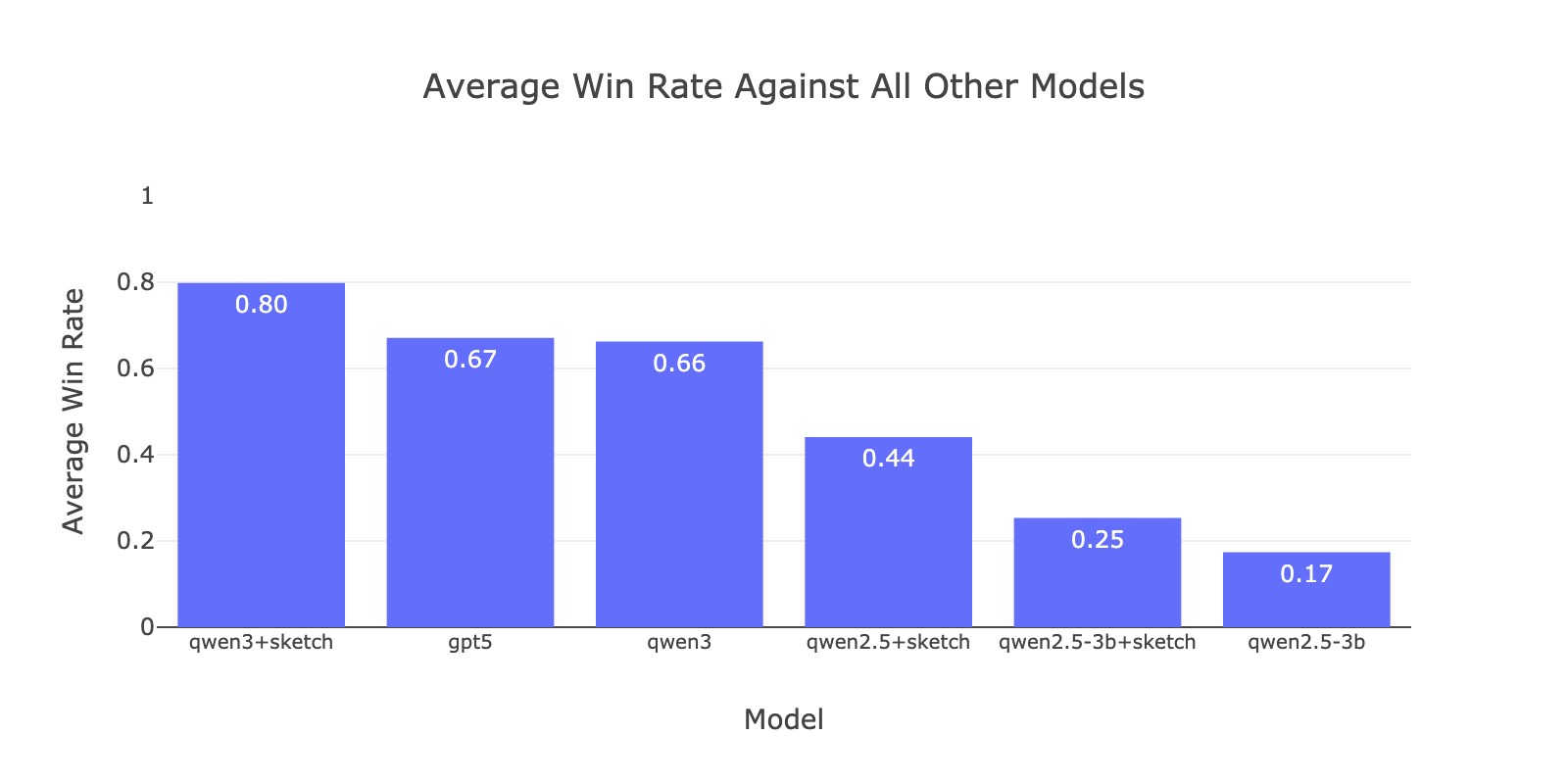}
        \Description{A bar chart that shows each model's average win rate against other models. The models from best to worst: qwen3+sketch, gpt5, qwen3, qwen2.5+sketch, qwen2.5-3b+sketch, qwen2.5-3b. Their win rates are 0.80, 0.67, 0.66, 0.44, 0.25, 0.17, respectively.}
    \end{subfigure}

    \caption{Rating scores (Top) and average win rate (Bottom) of models in our model generalization evaluation. 
    \textmd{We computed the rating scores by using the LMSYS calculation methodology~\cite{chiang2024chatbot,zheng2023chatbotarena}, and higher scores indicate models that were more often preferred by human judges. Bars show the median score and 95\% confidence intervals generated using bootstrap sampling.}}
    \label{fig:eloscores_comparison}
\end{figure}
Figure \ref{fig:eloscores_comparison} shows the results of our evaluation.
Among the tested models, Qwen3-Coder finetuned with our sketch reward model performed the best, while the base Qwen2.5-Coder 3B model performed the worst.
The additional base models included in this experiment performed roughly as we expected, where Qwen2.5-Coder 3B performed worse than its 32B counterpart \textcolor{black}{by a statistically significant margin} and the newer Qwen3-Coder performed better \textcolor{black}{by a statistically significant margin}.
Although the base Qwen3-Coder has roughly the same number of parameters (30B), it significantly outperformed the best-performing model from the feedback fine-tuning evaluation, Qwen2.5-Coder + Sketch.
The Qwen3 technical report documents a wide range of improvements to model architecture, training data, and training techniques~\cite{yang2025qwen3}.
For example, one improvement is that Qwen3 models were trained on roughly double (36 trillion tokens~\cite{yang2025qwen3}) the amount of overall data as Qwen2.5 models (18 trillion tokens~\cite{hui2024qwen2}).
This observation suggests that strategies for improving the general performance of LLMs can also lead to improvements in specific domains, like UI generation.

Nevertheless, we showed that our reward model trained with only 181 feedback samples consistently improved all tested base models \textcolor{black}{by a statistically significant amount}.
Notably, the best-performing Qwen3-Coder + Sketch model generated UI designs that were \textcolor{black}{significantly} preferred by judges over those from GPT-5, a proprietary LLM which has been estimated to be several orders of magnitude larger~\cite{albergotti2023smallmodels}.
This suggests that design expertise and specialized fine-tuning can greatly improve the efficiency of learning UI generation.

\begin{shaded}
Our results show that fine-tuning with our sketch reward model consistently led to improvements in UI generation capabilities for all tested baselines, suggesting generalizability.
We also show that a small amount of high-quality expert feedback can efficiently enable smaller models to outperform larger proprietary LLMs in UI generation.
\end{shaded}
\section{Discussion}
The results of our experiments show that UI code generation models can be improved with input from expert designers.
However, the method used to elicit designer feedback and convert it into machine-learnable data has a significant impact on the resulting model performance.
In this section, we discuss the implications of our work for data collection methodologies and learning from designer feedback. We conclude with limitations and avenues for future work.

\subsection{Agreeing to Disagree}
One challenge of our work and other human-centered problems is handling subjectivity and multiple resolutions of design problems.
Both phenomena can also lead to high variance in responses, which poses challenges for widely-used ranking feedback mechanisms.

This is especially true for UI design and design in general.
Our experimental results validate this in multiple ways.
First, in our UI preference pair quality assessment (Section~\ref{sec:data_noise_analysis}), six researchers ranked UI preference pairs and had very low levels of agreement on this ranking with expert designers, similar to findings by prior work~\cite{wu2024uiclip}.
While the speed of the ranking condition resulted in the largest volume of training data (1063 samples), this was not enough to offset the low data quality.
In our feedback fine-tuning evaluation, we found that fine-tuning an LLM on data generated from the ranking condition degraded the model's performance, resulting in the worst outcome. 

\textcolor{black}{It is possible to improve this existing ranking framework in several ways.}
\textcolor{black}{For example, previous work~\cite{gajos2005preference} has investigated uncertainty-based approaches to selecting binary pairs to query users, which may improve the sample efficiency of pairwise comparisons.}
\textcolor{black}{Another possibility is to develop more stringent rubrics to encourage more repeatable evaluation of UI screens~\cite{biyani2024rubicon}.}
However, our work advocates for a different approach: instead of presenting designers with a decision where disagreement can result in ``noise,'' (i.e., ranking UI screens) our approach can allow different designers to each give non-conflicting alternatives (i.e., UI design revisions) that are useful for representing for design assessment.
Suppose there is a UI screen where an important piece of information is not highlighted, e.g., a checkout screen whose information hierarchy makes it difficult to locate the total shopping cost.
Multiple designers might be able to diagnose the same problem but propose different fixes, e.g., highlighting the important element with color or moving it to a more isolated area to draw attention.
Whereas a ranking task would force the designer to choose between two model-generated UI screens, where the alternative possibly does not address the problem (or potentially even introduces new ones), our approach is able to introduce a new data point that specifically targets the original UI screen's flaws.
By ``agreeing to disagree,'' we train the model to consider multiple solutions for a problem, instead of a single canonical answer.
Anecdotally, informal conversations with designers revealed that they often generate and evaluate multiple alternative UI designs so incorporating multiple UI design revisions into model training aligns better with designer practice.

Other research in the machine learning literature has also explored feedback mechanisms outside of standard ranking, for example through online imitation learning of human demonstrations~\cite{shaikh2025aligning} or model alignment based on editing model-generated output~\cite{ji2024aligner}.
These approaches are motivated by the observation that desirable generations often lie outside a base model’s output distribution~\cite{shaikh2025aligning}, while conventional pairwise ranking limits feedback to in-distribution outputs.
This observation is especially relevant for UI generation, where excellent designs are often creative and uncommon, making them unlikely to emerge from a model’s typical outputs.
\subsection{Tradeoffs in Collecting Designer Feedback}
To capture feedback from expert designers, our paper explored tradeoffs in collecting and learning from designer feedback.

\textbf{Model-centric vs Designer-centric Interfaces. }
An early but important step is determining the requirements of feedback interfaces.
RLHF finetuning approaches (Section \ref{sec:background}) require datasets of ranked preference pairs to calibrate the probability of ``good'' and ``bad'' outputs, which motivates most conventional ranking interfaces.
This \textit{model-centric} approach to designing feedback interfaces primarily considers the requirements of machine learning algorithms; however, our work and previous work~\cite{wu2024uiclip} found that these feedback interfaces are often ineffective at capturing designer preference and critiques.
An alternative approach that we advance in this paper is to build interfaces around the existing practices and workflows of designers~\cite{hartmann2010d,o2018charrette}, which typically focus on improving a single UI screen in a non-comparative setting.
This \textit{designer-centric} approach for data collection involves converting artifacts from designers' day-to-day activities, such as design reviews, white-boarding, and direct edits, into machine-learnable data.
Our UI preference pair quality and model evaluation experiments indicate that this approach is effective in some contexts (e.g., sketching and revision) but not in others (e.g., commenting), suggesting the need to better understand and balance the goals of designers and model training.

\textbf{Quantity vs Quality.}
Our work investigates the tradeoff between data quantity and quality for training UI generation models.
Previously, the effectiveness of ML annotation systems (\textit{e.g.,} feedback interfaces) has often been measured by the speed at which they could produce responses from human annotators~\cite{little2010turkit,von2006games}.
For example, one proposed metric for evaluating annotation interfaces is the number of training examples that could be collected through the interaction in a given time-frame~\cite{von2006games}.
While these measurements may be more reflective of utility for tasks where voting~\cite{little2010turkit} or other forms of redundant checking~\cite{von2006games,von2008recaptcha} is possible, we found that they were not useful for UI design feedback interfaces.
In our work, we found that while some interactions require more time and effort to complete (\textit{e.g.,} design revision), they often result in better data quality.
Our approach to evaluating UI designs involved collecting feedback across multiple UI design feedback interfaces to train UI generation models and evaluate those models with human judges.
\textcolor{black}{The results of our data quality showed that sketching exhibited the best performance and balance between data quality and quantity. We note that these findings are valid for our chosen model configuration, and newer, stronger LLMs may have different performance characteristics, \textit{e.g.,} may have better ability to comprehend and apply comment-based fixes.
}
\textcolor{black}{Overall, the development of more standardized and easily calculable heuristics for UI design evaluation at an earlier stage remains an important direction for future work, especially to inform the earlier stages of design.}

\subsection{Limitations \& Future Work}
In this paper, we showed that our data collection and modeling approach leads to significant improvements to the quality of generated UI code by learning from designer feedback.
We see multiple avenues for improvement for our work.

First, our current work provides limited validation of our data collection approach.
Our designer feedback evaluation was relatively short (one hour), and designers evaluated synthetically-generated UIs, which are typically of lower quality than professional designs.
Our evaluation design was motivated by several practical factors (e.g., our desire to control conditions) and data confidentiality (designers may normally work on proprietary, unreleased designs).
\textcolor{black}{Our final model evaluation arenas were also limited to six judges, although we still found this number sufficient for statistically significant findings.}
\textcolor{black}{Future studies could investigate the use of more realistic training data consisting of professional designs to build a stronger reward model, a larger number of evaluators, and focus on investigating the real-world feasibility of our approach.}

We see numerous opportunities to collect other types of designer feedback and developing machine learning approaches to learn from them.
When asked about other types of feedback common in their work,\footnote{The question was aimed at understanding different mechanisms for recording feedback; however, many participants responded with methodologies of collecting it.} designers in our evaluation often referenced numerous methodologies in the HCI literature, such as usability studies, cognitive walkthroughs, and analyzing user engagement metrics.
The techniques we used to collect feedback would likely need to be adapted to infer complex interactive properties related to usability and accessibility.

Finally, the machine learning formulation to learn from these other types of design feedback data would need to be updated as well to translate these artifacts (e.g., recordings or transcripts) into machine-learnable datasets.
We designed our training setup for a single-screen evaluation.
For example, we employ a visual language model (i.e., CLIP) as a reward model to score a single image input.
Evaluating higher-level aspects of UX and interaction design would necessitate a more complex evaluator module that could potentially crawl and interact with an running version of the application (e.g., a UI agent).
The generator module must also learn from this feedback and use it to generate functionality for app navigation and handling interaction (e.g., a login flow).
We expect numerous opportunities for incorporating design expertise in advancements in LLM model training and their HCI applications.
\section{Conclusion}
In this paper, we presented interactions for collecting designer feedback based on common workflows, such as commenting, sketching, and revising. We introduce methods for converting this designer feedback into machine learnable datasets for UI assessment and generation.
We conducted a designer feedback evaluation where we asked twenty one designers to give feedback on a set of synthetically generated UIs and developed techniques to generate a designer feedback dataset from them. 
We analyzed this data to validate the quality of its labels, showing much higher levels of label agreement than existing ranking methods for output comparison.
To validate the usefulness of our data, we finetuned UI code generation models in a RLHF configuration with a reward model trained from different forms of designer feedback.
We conducted two arena-style model evaluations comparing the performance of our models and baselines through repeated blind evaluations of the generated UI design outputs.
Our results showed that feedback data collected from natural designer interactions led to better model performance compared to ranking data, which is widely used for model training.
In addition, our best-performing model outperformed all test baselines, including a larger proprietary model, highlighting the impact of a small amount of high-quality expert feedback.
Our work suggests applying designer-aligned interaction techniques is beneficial for training models to generate UIs.

\bibliographystyle{ACM-Reference-Format}
\bibliography{sample-base}

\appendix
\newpage
\section{Model Hyperparameters}
We provide all hyperparameters used for our model training experiments in Table \ref{tab:model_hyperparams}.
\begin{table*}[!htb] 
\caption{Hyperparameters used for our modeling training experiments.}
\begin{tabular}{@{}lll@{}} 
\toprule
Algorithm & Hyperparameters           & Value \\ \midrule
Reward Models  & max optim steps      & 100   \\
          & batch size           & 32    \\
          & weight decay         & 0.2   \\
          & learning rate        & 1e-3  \\
          & margin               & 1e-2  \\
          & UIClip aug prob      & 0.5   \\
Generator Models & odds ratio weight    & 1.0   \\
          & effective batch size & 8     \\
          & context length       & 4096  \\
          & learning rate        & 5e-6  \\ \bottomrule
\end{tabular}\Description{Table of hyperparameters used in our experiment. The reward model training used six hyperparameters, max optim steps, batch size, weight decay, learning rate, margin, and UIClip aug prob. Their values are 100, 32, 0.2, 1e-3, 1e-2, and 0.5, respectively. The generator model training used four hyperparameters, odds ratio weight, effective batch size, context length, and learning rate. Their values are 1.0, 8, 4096, and 5e-6, respectively.}
\label{tab:model_hyperparams}
\end{table*}
\section{Model Prompts}
We used several prompts in our experiments to train and run large language models.

We used the the following prompt for generating HTML web pages from a short natural language description. We also fine-tuned our model using prompts formatted in this template. This prompt was determined by trial and error and manual inspection on a small number of test cases.

\begin{lstlisting}
provide the complete HTML code for a web page implemented with only tailwind CSS and font awesome icons. do not use any templating languages like jinja. the result should resemble an award-winning iOS app. include realistic and complete placeholder data. do not treat this as the starting point for an app - it should be the mockup of a final complete UI. remember to include alt text for all images. do not use javascript. do not use SVGs. here is a description of the webpage: <natural language description>
\end{lstlisting}

We used several prompts for evaluating screenshots. Our approach was to compute an augmented text embedding using a mix of positive and negative prompts.

We used the original prompt format described in UIClip~\cite{wu2024uiclip} as the positive prompt.

\begin{lstlisting}
ui screenshot. well-designed. <natural language description>
\end{lstlisting}

We also computed a negative prompt to represent a poorly designed version of the screen.

\begin{lstlisting}
ui screenshot. poor design. <natural language description>
\end{lstlisting}

Finally, we computed a second negative prompt to represent empty or overly simple UIs.

\begin{lstlisting}
ui screenshot. poor design. empty screen
\end{lstlisting}

Text embeddings were computed for each of these prompts and combined together using the following equation.

\begin{equation}
    \mathbf{v}^* = \mathbf{v}_{pos} - 0.5 \cdot ( 0.9 \cdot \mathbf{v}_{\text{neg}} + 0.1 \cdot \mathbf{v}_{\text{empty}} )
\end{equation}

$\mathbf{v}^*$ refers to the final text embedding used for UIClip score calculation.
$\mathbf{v}_{pos}$ refers to the positive text embedding.
$\mathbf{v}_{neg}$ refers to the negative text embedding.
$\mathbf{v}_{empty}$ refers to the empty text embedding.
The weighting between the positive and negative embeddings were determined by trial and error and manual inspection on a small number of UI screenshots.

We prompted Qwen2.5-Coder 32B to improve UIs given a list of designer-authored comments.

\begin{lstlisting}
i have implemented a website using only html, tailwind css, and font awesome icons.

```html
<original UI HTML code>
```

a designer has wrote some notes and feedback:
"<list of comments>"

incorporate this feedback into the website code. you must respond with the entire code implementation. do not use comments that are placeholders for the original code.
\end{lstlisting}

We used a similar prompt to improve UIs given a list of region-grounded annotations provided by designers.

\begin{lstlisting}
i have implemented a website using only html, tailwind css, and font awesome icons.

```html
<original UI HTML code>
```

a designer has wrote some notes and feedback for several regions of the HTML:
"<list of comments paired with HTML snippets>"

incorporate this feedback into the website code. you must respond with the entire code implementation. do not use comments that are placeholders for the original code.
\end{lstlisting}
\section{Example Outputs} \label{sec:example_outputs}
\begin{figure*}[!htb]
    \centering
    \includegraphics[width=0.8\linewidth]{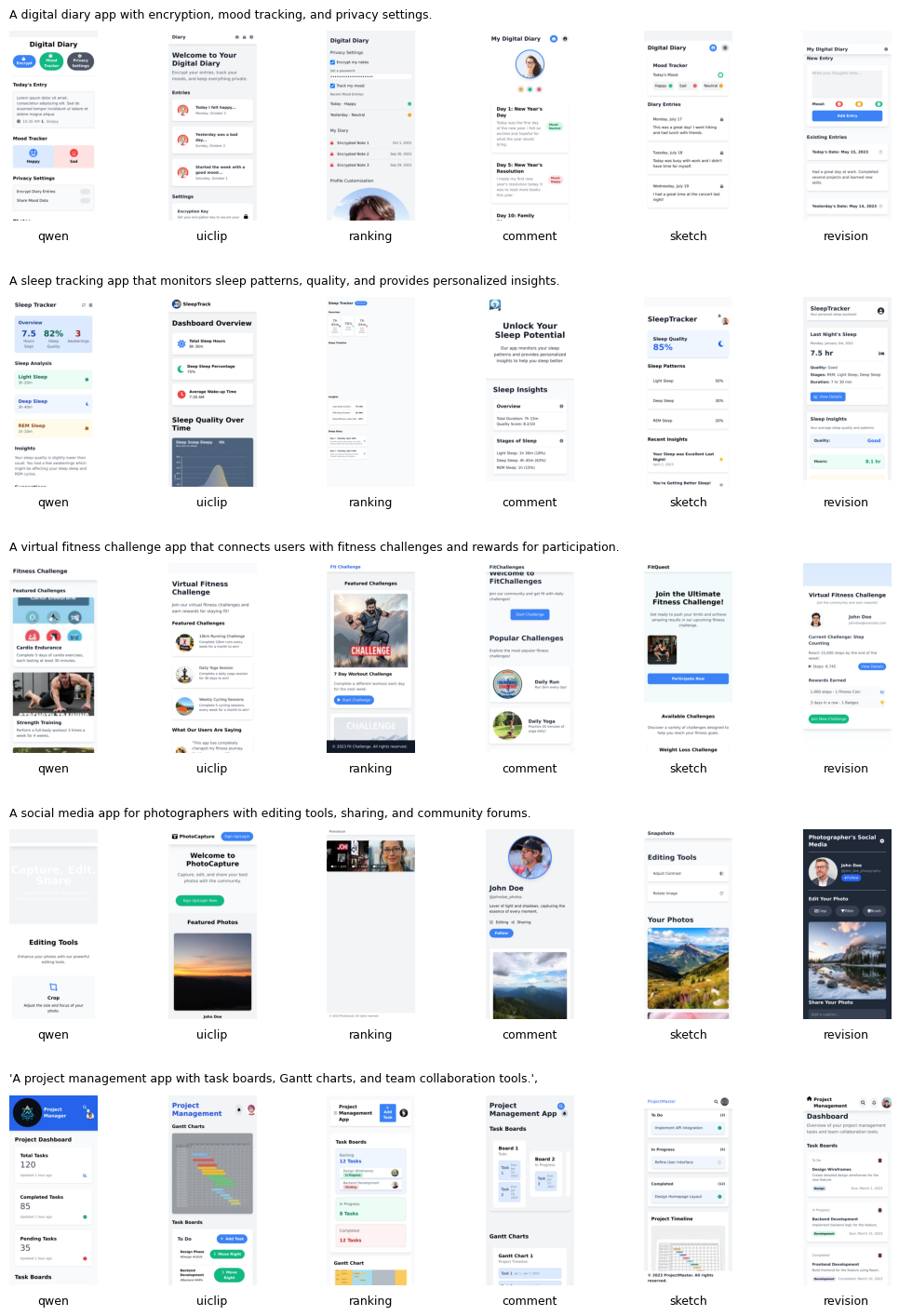}
    \Description{A grid of six by five UI screenshots that shows the rendered output of six tested models for five randomly sampled text descriptions from our feedback fine-tuning evaluation. In general, all models produce plausible outputs but also make varying numbers of design flaws.}
    \caption{Figure shows rendered output of six models tested in the feedback fine-tuning evaluation. We rendered model outputs for five randomly sampled text descriptions from our evaluation set.}
    \label{fig:example-output}
\end{figure*}

We provide some example outputs from our model evaluation experiment in Figure \ref{fig:example-output}.

\begin{figure*}[!htb]
    \centering
    \includegraphics[width=0.8\linewidth]{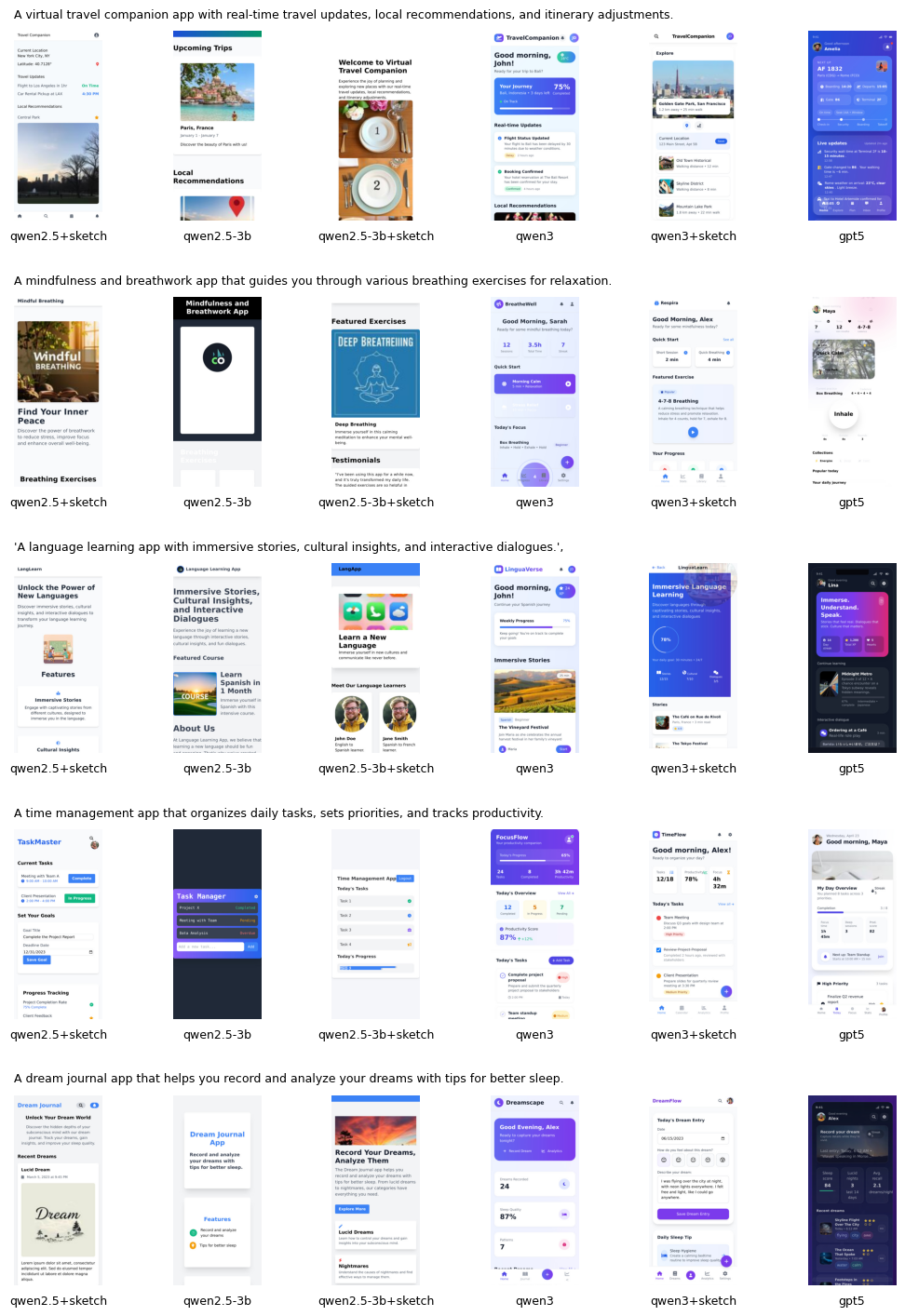}
    \Description{A grid of six by five UI screenshots that shows the rendered output of eight tested models for five randomly sampled text descriptions from our model generalization evaluation. In general, all models produce plausible outputs but also make varying numbers of design flaws.}
    \caption{Figure shows rendered output of six models tested in the model generalization evaluation. We rendered model outputs for five randomly sampled text descriptions from our evaluation set.}
    \label{fig:example-output2}
\end{figure*}
\section{Statistical Significance Analysis}\label{sec:wald_test}
\begin{figure*}[t]
    \centering
    \begin{subfigure}{0.48\linewidth}
        \centering
        \includegraphics[width=\linewidth]{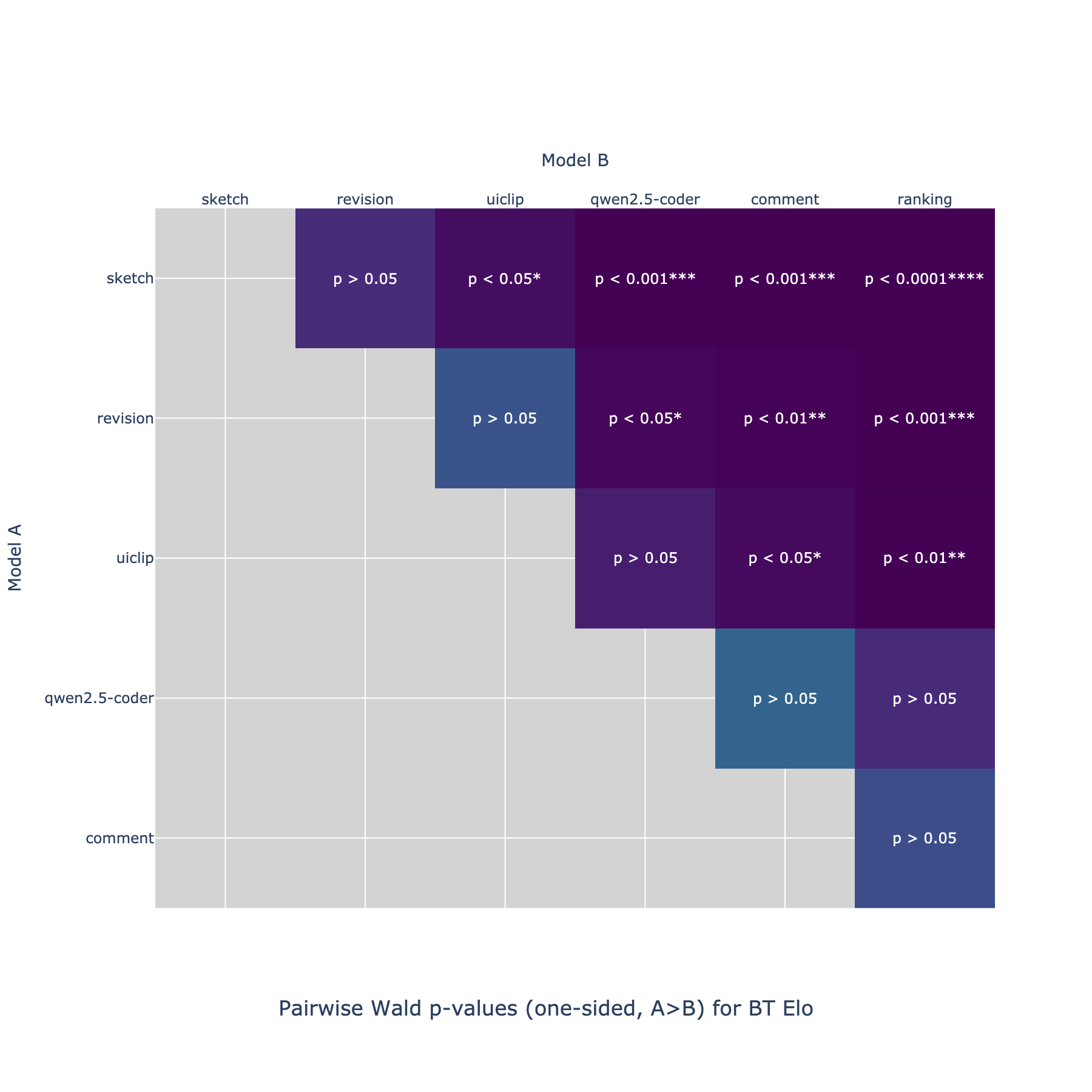}
        \Description{A matrix where the axes display models tested in the feedback fine-tuning evaluation, sorted by their Elo rating. The intersecting cell contains the p-value for the hypothesis that model A is better than model B. The sketch model is significantly better than all other models besides revision. The revision model is significantly better than default qwen, the comment model, and ranking model. The uiclip-tuned model is significantly better than the comment and ranking models.}
    \end{subfigure}
    \hfill
    \begin{subfigure}{0.48\linewidth}
        \centering
        \includegraphics[width=\linewidth]{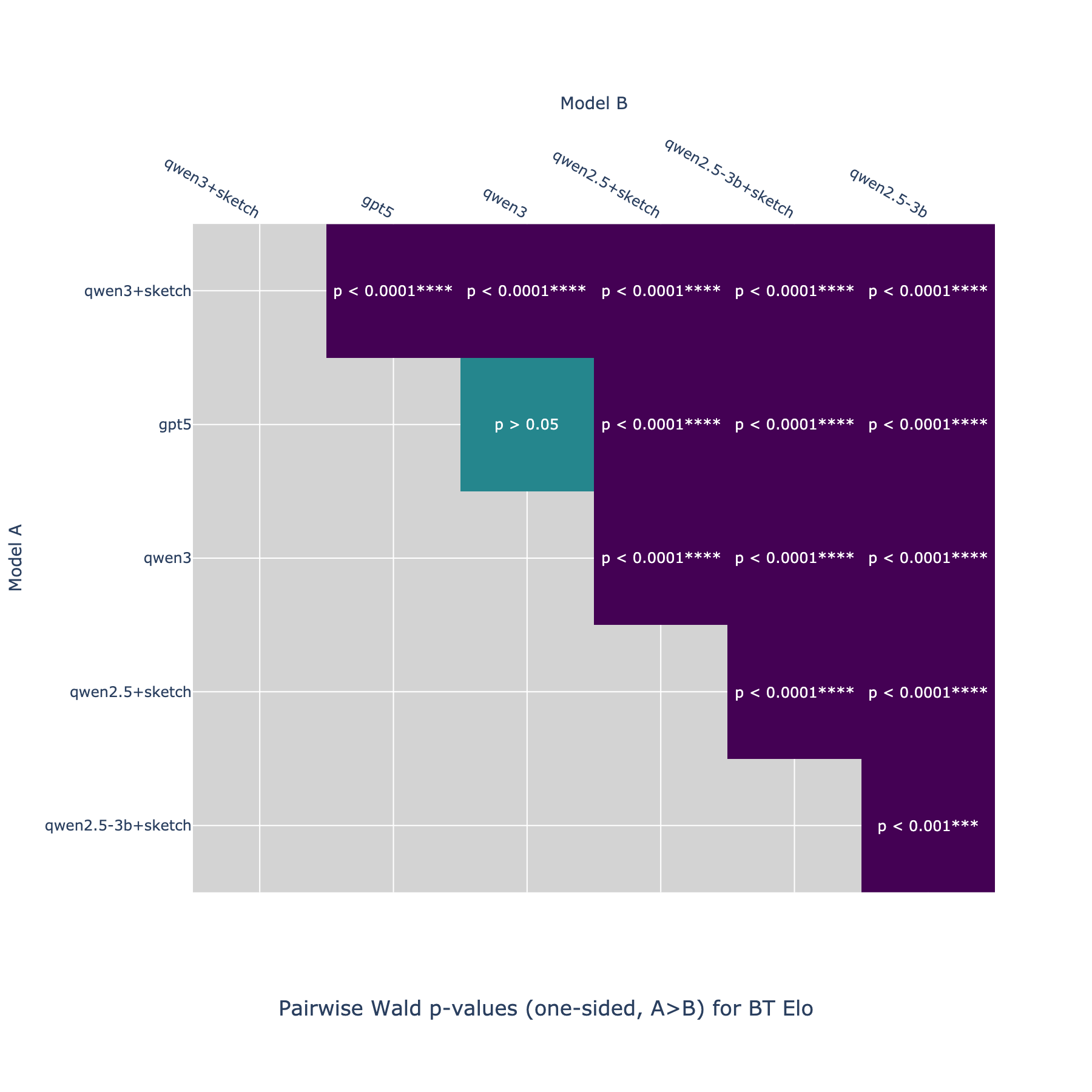}
        \Description{A matrix where the axes display models tested in the model generalization evaluation, sorted by their Elo rating. The intersecting cell contains the p-value for the hypothesis that model A is better than model B. The qwen3+sketch model is significantly better than all other models. GPT-5 is significantly better than qwen2.5+sketch, qwen2.5-3b+sketch, and qwen2.5-3b. Qwen3 is significantly better than qwen2.5+sketch, qwen2.5-3b+sketch, and qwen2.5-3b. Qwen2.5+sketch is significantly better than qwen2.5-3b+sketch and qwen2.5-3b. Qwen2.5-3b+sketch is significantly better than qwen2.5-3b.}
    \end{subfigure}

    \caption{\textcolor{black}{Matrices show the p-values for model comparisons in the Feedback Comparison (Left) and Model Generalization (Right) studies.
    \textmd{The axes display models tested in the each evaluation, sorted by their Elo rating. The intersecting cell contains the one-sided p-value for the hypothesis that model A is better than model B. We consider only pairs where the rating of model A is higher than that of model B}}}
    \label{fig:pvalue_matrix}
\end{figure*}

\textcolor{black}{To complement the 95\% confidence interval produced by bootstrap analysis~\cite{chiang2024chatbot}, we conducted a Wald test~\cite{wald1943tests} to test the significance of model performance differences predicted Elo ratings. We compare each model pair from our feedback comparison and model generalization studies where the rating of model A is greater than model B. The results are shown in the matrix Figure \ref{fig:pvalue_matrix}.}
\end{document}